\documentclass[twocolumn,floatfix,aps,superscriptaddress]{revtex4}
\usepackage{graphicx}
\usepackage{amsmath}
\usepackage{amssymb}
\usepackage{bm}
\usepackage{hyperref}

\usepackage{color}

\begin{document}

\title{Valley-momentum locking in a graphene superlattice\\
with Y-shaped Kekul\'{e} bond texture}
\author{O. V. Gamayun}
\affiliation{Instituut-Lorentz, Universiteit Leiden, P.O. Box 9506, 2300 RA Leiden, The Netherlands}
\author{V. P. Ostroukh}
\affiliation{Instituut-Lorentz, Universiteit Leiden, P.O. Box 9506, 2300 RA Leiden, The Netherlands}
\author{N. V. Gnezdilov}
\affiliation{Instituut-Lorentz, Universiteit Leiden, P.O. Box 9506, 2300 RA Leiden, The Netherlands}
\author{\.{I}. Adagideli}
\affiliation{Faculty of Engineering and Natural Sciences, Sabanci University, Orhanli-Tuzla, 34956 Istanbul, Turkey}
\author{C. W. J. Beenakker}
\affiliation{Instituut-Lorentz, Universiteit Leiden, P.O. Box 9506, 2300 RA Leiden, The Netherlands}

\date{October 2017}
\begin{abstract}
Recent experiments by Guti\'{e}rrez \textit{et al.}\ [Nature Phys.\ \textbf{12}, 950 (2016)] on a graphene-copper superlattice have revealed an unusual Kekul\'{e} bond texture in the honeycomb lattice --- a Y-shaped modulation of weak and strong bonds with a wave vector connecting two Dirac points. We show that this socalled ``Kek-Y'' texture produces two species of massless Dirac fermions, with valley isospin locked parallel or antiparallel to the direction of motion. In a magnetic field $B$ the valley degeneracy of the $B$-dependent Landau levels is removed by the valley-momentum locking --- but a $B$-independent and valley-degenerate zero-mode remains.  
\end{abstract}
\maketitle

\section{Introduction}
\label{intro}

The coupling of orbital and spin degrees of freedom is a promising new direction in nano-electronics, referred to as ``spin-orbitronics'', that aims at non-magnetic control of information carried by charge-neutral spin currents \cite{Fer08,Aws09,Kus15}. Graphene offers a rich platform for this research \cite{Rec11,Pes12}, because the conduction electrons have three distinct spin quantum numbers: In addition to the spin magnetic moment $s=\pm 1/2$, there is the sublattice pseudospin $\sigma=\text{A,B}$ and the valley isospin $\tau=K,K'$. While the coupling of the electron spin $s$ to its momentum $p$ is a relativistic effect, and very weak in graphene, the coupling of $\sigma$ to $p$ is so strong that one has a pseudospin-momentum locking: The pseudospin points in the direction of motion, as a result of the helicity operator $\bm{p}\cdot\bm{\sigma}\equiv p_x\sigma_x+p_y\sigma_y$ in the Dirac Hamiltonian of graphene.

\begin{figure}[tb]
\centerline{\includegraphics[width=0.9\linewidth]{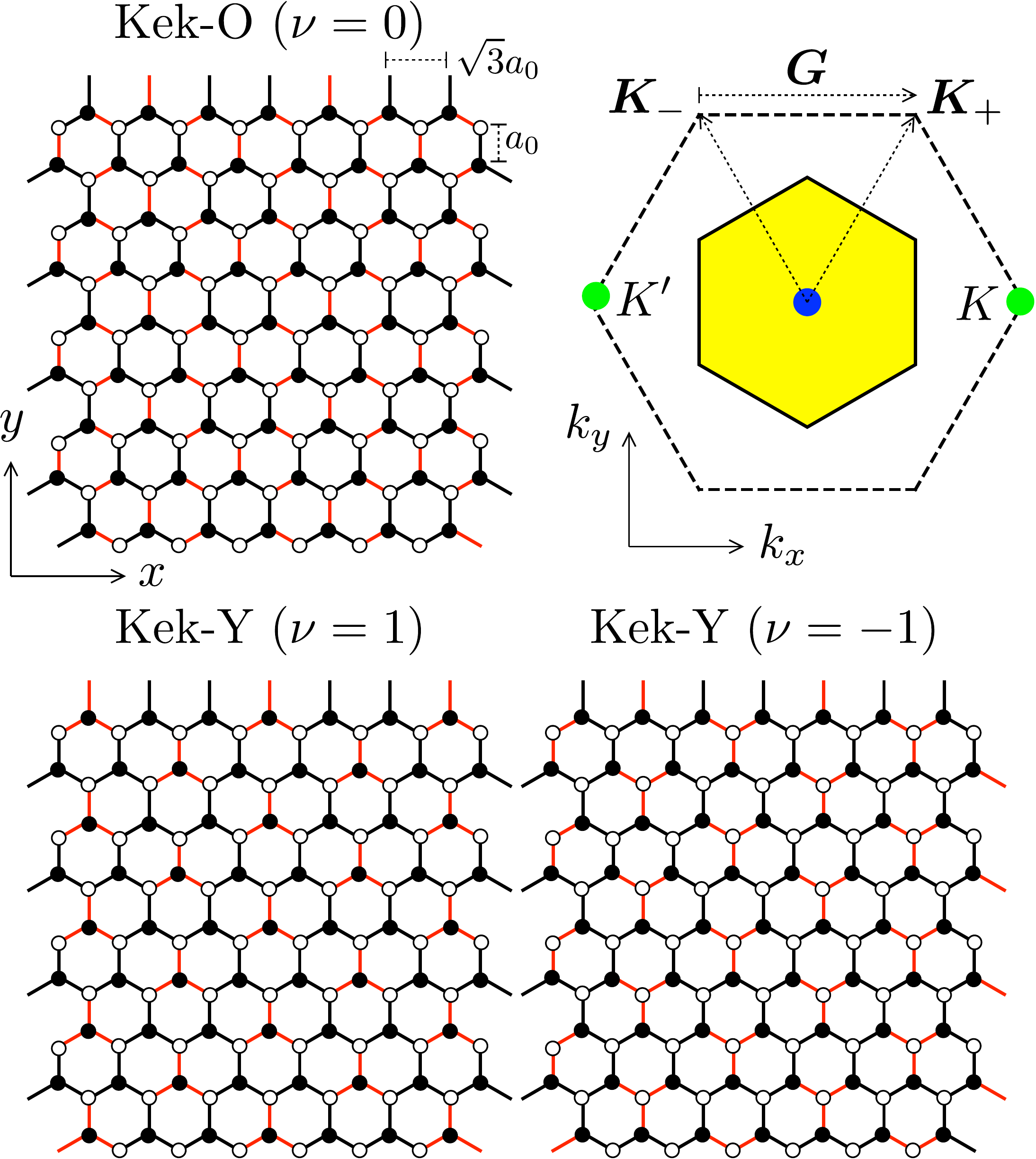}}
\caption{Honeycomb lattices with a Kek-O or Kek-Y bond texture, all three sharing the same superlattice Brillouin zone (yellow hexagon, with reciprocal lattice vectors $\bm{K}_\pm$). Black and white dots label A and B sublattices, black and red lines distinguish different bond strengths. The lattices are parametrized according to Eq.\ \eqref{trndef} (with $\phi=0$) and distinguished by the index $\nu=1+q-p$ modulo $3$ as indicated. The $K$ and $K'$ valleys (at the green Dirac points) are coupled by the wave vector $\bm{G}=\bm{K}_+-\bm{K}_-$ of the Kekul\'{e} bond texture and folded onto the center of the superlattice Brillouin zone (blue point). 
}
\label{FLattice}
\end{figure}

The purpose of this paper is to propose a way to obtain a similar handle on the valley isospin, by adding a term $\bm{p}\cdot\bm{\tau}$ to the Dirac Hamiltonian, which commutes with the pseudospin helicity and locks the valley to the direction of motion. We find that this valley-momentum locking should appear in a superlattice that has recently been realized experimentally by Guti\'{e}rrez \textit{et al.}\ \cite{Gut16,Gut15}: A superlattice of graphene grown epitaxially onto Cu(111), with the copper atoms in registry with the carbon atoms. One of six carbon atoms in each superlattice unit cell ($\sqrt 3\times\sqrt 3$ larger than the original graphene unit cell) have no copper atoms below them and acquire a shorter nearest-neighbor bond. The resulting Y-shaped periodic alternation of weak and strong bonds (see Fig.\ \ref{FLattice}) is called a Kekul\'{e}-Y (Kek-Y) ordering, with reference to the Kekul\'{e} dimerization in a benzene ring (called Kek-O in this context) \cite{Gut15}.

The Kek-O and KeK-Y superlattices have the same Brillouin zone, with the $K$ and $K'$ valleys of graphene folded on top of each other. The Kek-O ordering couples the valleys by opening a gap in the Dirac cone \cite{Cha00,Hou07,Che09a,Che09b,Gom12}, and it was assumed by Guti\'{e}rrez \textit{et al.}\ that the same applies to the Kek-Y ordering \cite{Gut16,Gut15}. While it is certainly possible that the graphene layer in the experiment is gapped by the epitaxial substrate (for example, by a sublattice-symmetry breaking ionic potential \cite{Gio07,Gio15,Ren15}), we find that the Y-shaped Kekul\'{e} bond ordering by itself does not impose a mass on the Dirac fermions \cite{note5}. Instead, the valley degeneracy is broken by the helicity operator $\bm{p}\cdot\bm{\tau}$, which preserves the gapless Dirac point while locking the valley degree of freedom to the momentum. In a magnetic field the valley-momentum locking splits all Landau levels except for the zeroth Landau level, which remains pinned to zero energy. 

\section{Tight-binding model}
\label{TBmodel}

\subsection{Real-space formulation}
\label{TBreal}

A monolayer of carbon atoms has the tight-binding Hamiltonian
\begin{equation}
H = - \textstyle{\sum_{\bm{r}}\sum_{\ell=1}^3}  t_{\bm{r},\ell}\,a_{\bm{r}}^\dagger b^{\vphantom{\dagger}}_{\bm{r}+\bm{s}_{\ell}} + {\rm H.c.},\label{Ham1}
\end{equation}
describing the hopping with amplitude $t_{\bm{r},\ell}$ between an atom at site $\bm{r}=n\bm{a}_1+m\bm{a}_2$ ($n,m\in\mathbb{Z}$) on the A sublattice (annihilation operator $a_{\bm{r}}$) and each of its three nearest neighbors at $\bm{r}+\bm{s}_\ell$ on the B sublattice (annihilation operator $b_{\bm{r}+\bm{s}_\ell}$). The lattice vectors are defined by $\bm{s}_1=\tfrac{1}{2}(\sqrt{3},-1)$, $\bm{s}_2=-\tfrac{1}{2}(\sqrt{3},1)$, $\bm{s}_3=(0,1)$, $\bm{a}_1=\bm{s}_3-\bm{s}_1$,  $\bm{a}_2=\bm{s}_3-\bm{s}_2$. All lengths are measured in units of the unperturbed C--C bond length $a_0\equiv 1$.

For the uniform lattice, with $t_{\bm{r},\ell}\equiv t_0$, the band structure is given by \cite{Cas09}
\begin{equation}
E(\bm{k}) = \pm |\varepsilon(\bm{k})|,\;\; \varepsilon(\bm{k})=t_0\textstyle{\sum_{\ell=1}^3} e^{i \bm{k}\cdot\bm{s}_\ell}.\label{Eunperturbed}
\end{equation}
There is a conical singularity at the Dirac points $\bm{K}_{\pm} = \tfrac{2}{9}\pi\sqrt{3}(\pm 1,\sqrt{3})$, where $E (\bm{K}_{\pm})=0$. For later use we note the identities
\begin{equation}
\varepsilon(\bm{k})=\varepsilon(\bm{k}+3\bm{K}_\pm)=e^{ 2\pi i/3}\varepsilon(\bm{k}+ \bm{K}_+ + \bm{K}_-).\label{epsilonidentities}
\end{equation}

The bond-density wave that describes the Kek-O and Kek-Y textures has the form
\begin{subequations}
\label{trndef}
\begin{align}
t_{\bm{r},\ell}/t_0 &=1+2\,{\rm Re}\,\bigl[\Delta e^{i (p\bm{K}_+ +q\bm{K}_-)\cdot\bm{s}_\ell+i\bm{G}\cdot\bm{r}}\bigr]\label{trndefa}\\
&=1+2\Delta_0\cos[\phi+\tfrac{2}{3}\pi(m-n+N_\ell)],\label{trndefb}\\
&\quad N_1=-q,\;\;N_2=-p,\;\;N_3=p+q,\;\;p,q\in\mathbb{Z}_3.\nonumber
\end{align}
\end{subequations}
The Kekul\'{e} wave vector
\begin{equation}
\bm{G}\equiv \bm{K}_+ - \bm{K}_-=\tfrac{4}{9}\pi\sqrt{3}( 1,0)\label{Gdef}
\end{equation}
couples the Dirac points. The coupling amplitude $\Delta=\Delta_0 e^{i\phi}$ may be complex, but the hopping amplitudes $t_{\bm{r},\ell}$ are real in order to preserve time-reversal symmetry.

As illustrated in Fig.\ \ref{FLattice}, the index 
\begin{equation}
\nu=1+q-p\mod 3\label{nudef}
\end{equation}
distinguishes the Kek-O texture ($\nu=0$) from the Kek-Y texture ($\nu=\pm 1$). Each Kekul\'{e} superlattice has a $2\pi/3$ rotational symmetry, reduced from the $2\pi/6$ symmetry of the graphene lattice. The two $\nu=\pm 1$ Kek-Y textures are each others mirror image \cite{note0}.
 
\subsection{Transformation to momentum space}
\label{fromrealtomomentum}

The Kek-O and Kek-Y superlattices have the same hexagonal Brillouin zone, with reciprocal lattice vectors $\bm{K}_\pm$ --- smaller by a factor $1/\sqrt 3$ and rotated over $30^\circ$ with respect to the original Brillouin zone of graphene (see Fig.\ \ref{FLattice}). The Dirac points of unperturbed graphene are folded from the corner to the center of the Brillouin zone and coupled by the bond density wave. 

To study the coupling we Fourier transform the tight-binding Hamilonian \eqref{Ham1},
\begin{align}
H(\bm{k}) ={}& - \varepsilon(\bm{k})a_{\bm k}^\dagger b^{\vphantom{\dagger}}_{\bm k} - \Delta  \varepsilon(\bm{k}+p\bm{K}_+ +q\bm{K}_-)a^\dagger_{\bm{k}+\bm{G}} b^{\vphantom{\dagger}}_{\bm k}\nonumber\\
&- \Delta^\ast  \varepsilon(\bm{k}-p\bm{K}_+ -q\bm{K}_-) a_{\bm{k}-\bm{G}}^\dagger b^{\vphantom{\dagger}}_{\bm k}+ {\rm H.c.}\label{Hkdef}
\end{align}
The momentum $\bm{k}$ still varies over the original Brillouin zone. In order to restrict it to the superlattice Brillouin zone we collect the annihilation operators at $\bm{k}$ and $\bm{k}\pm\bm{G}$ in the column vector $c_{\bm k}=(a_{\bm k},a_{\bm{k}-\bm{G}},a_{\bm{k}+\bm{G}},b_{\bm k},b_{\bm{k}-\bm{G}},b_{\bm{k}+\bm{G}})$ and write the Hamiltonian in a $6\times 6$ matrix form:
\begin{subequations}
\label{Hkreduceddef}
\begin{align}
&H(\bm{k})=-c^\dagger_{\bm{k}}\begin{pmatrix}
0&{\cal E}_\nu(\bm{k})\\
{\cal E}^\dagger_\nu(\bm{k})&0
\end{pmatrix}c_{\bm k},\\
&{\cal E}_\nu=\begin{pmatrix}
\varepsilon_0&\tilde{\Delta}\varepsilon_{\nu+1}&\tilde{\Delta}^\ast\varepsilon_{-\nu-1}\\
\tilde{\Delta}^\ast\varepsilon_{1-\nu}&\varepsilon_{-1}&\tilde{\Delta}\varepsilon_{\nu}\\
\tilde{\Delta}\varepsilon_{\nu-1}&\tilde{\Delta}^\ast\varepsilon_{-\nu}&\varepsilon_1
\end{pmatrix},\\
&\tilde{\Delta}=e^{2\pi i(p+q)/3}\Delta,\;\;\varepsilon_n=\varepsilon(\bm{k}+n\bm{G}),
\end{align}
\end{subequations}
where we used Eq.\ \eqref{epsilonidentities}.

\section{Low-energy Hamiltonian}
\label{lowEH}

\subsection{Gapless spectrum}
\label{gaplessspectrum}

The low-energy spectrum is governed by the four modes $u_{\bm k}=(a_{\bm{k}-\bm{G}},a_{\bm{k}+\bm{G}},b_{\bm{k}-\bm{G}},b_{\bm{k}+\bm{G}})$, which for small $\bm{k}$ lie near the Dirac points at $\pm \bm{G}$. (We identify the $K$ valley with $+\bm{G}$ and the $K'$ valley with $-\bm{G}$.) Projection onto this subspace reduces the six-band Hamiltonian \eqref{Hkreduceddef} to an effective four-band Hamiltonian,
\begin{equation}
H_{\rm eff}=-u^\dagger_{\bm{k}}\begin{pmatrix}
0&h_\nu\\
h^\dagger_\nu&0
\end{pmatrix}u_{\bm k},\;\;
h_\nu=\begin{pmatrix}
\varepsilon_{-1}&\tilde{\Delta}\varepsilon_{\nu}\\
\tilde{\Delta}^\ast\varepsilon_{-\nu}&\varepsilon_1
\end{pmatrix}.\label{Hkeff}
\end{equation}
Corrections to the low-energy spectrum from virtual transitions to the higher bands are of order $\Delta_0^2$. We will include these corrections later, but for now assume $\Delta_0\ll 1$ and neglect them. 

The $\bm{k}$-dependence of $\varepsilon_n$ may be linearized near $\bm{k}=0$,
\begin{equation}
\varepsilon_0=3t_0,\;\;\varepsilon_{\pm 1}=\hbar v_0(\mp k_x +ik_y)+\text{order}\,(k^2),\label{vFdef}
\end{equation}
with Fermi velocity $v_0=\tfrac{3}{2}t_0 a_0/\hbar$. The corresponding 4-component Dirac equation has the form
\begin{subequations}
\label{HDiracdef}
\begin{align}
&{\cal H}
\begin{pmatrix}
\Psi_{K'}\\
\Psi_{K}
\end{pmatrix}=E\begin{pmatrix}
\Psi_{K'}\\
\Psi_{K}
\end{pmatrix},\;\;
{\cal H}=\begin{pmatrix}
v_0\bm{p}\cdot\bm{\sigma}&\tilde{\Delta}Q_\nu\\
\tilde{\Delta}^\ast Q_\nu^\dagger&v_0\bm{p}\cdot\bm{\sigma}
\end{pmatrix},\label{HDiracdefa}
\\
&
\Psi_{K'}=\begin{pmatrix}
-\psi_{B,K'}\\
\psi_{A,K'}
\end{pmatrix},\;\;
\Psi_{K}=\begin{pmatrix}
\psi_{A,K}\\
\psi_{B,K}
\end{pmatrix},
\label{HDiracdefb}
\\
&Q_\nu=\begin{pmatrix}
\varepsilon_{-\nu}^\ast&0\\
0&-\varepsilon_{\nu}
\end{pmatrix}=\begin{cases}
3t_0\sigma_z&\text{if}\;\;\nu=0,\\
v_0(\nu p_x-ip_y)\sigma_0&\text{if}\;\;|\nu|=1.
\end{cases}\label{HDiracdefc}
\end{align}
\end{subequations}
The spinor $\Psi_{K}$ contains the wave amplitudes on the $A$ and $B$ sublattices in valley $K$ and similarly $\Psi_{K'}$ for valley $K'$, but note the different ordering of the components \cite{note1}. We have defined the momentum operator $\bm{p}=-i\hbar\partial/\partial\bm{r}$, with $\bm{p}\cdot\bm{\sigma}=p_x\sigma_x+p_y\sigma_y$. The Pauli matrices $\sigma_x,\sigma_y,\sigma_z$, with $\sigma_0$ the unit matrix, act on the sublattice degree of freedom.

\begin{figure}[tb]
\centerline{\includegraphics[width=0.7\linewidth]{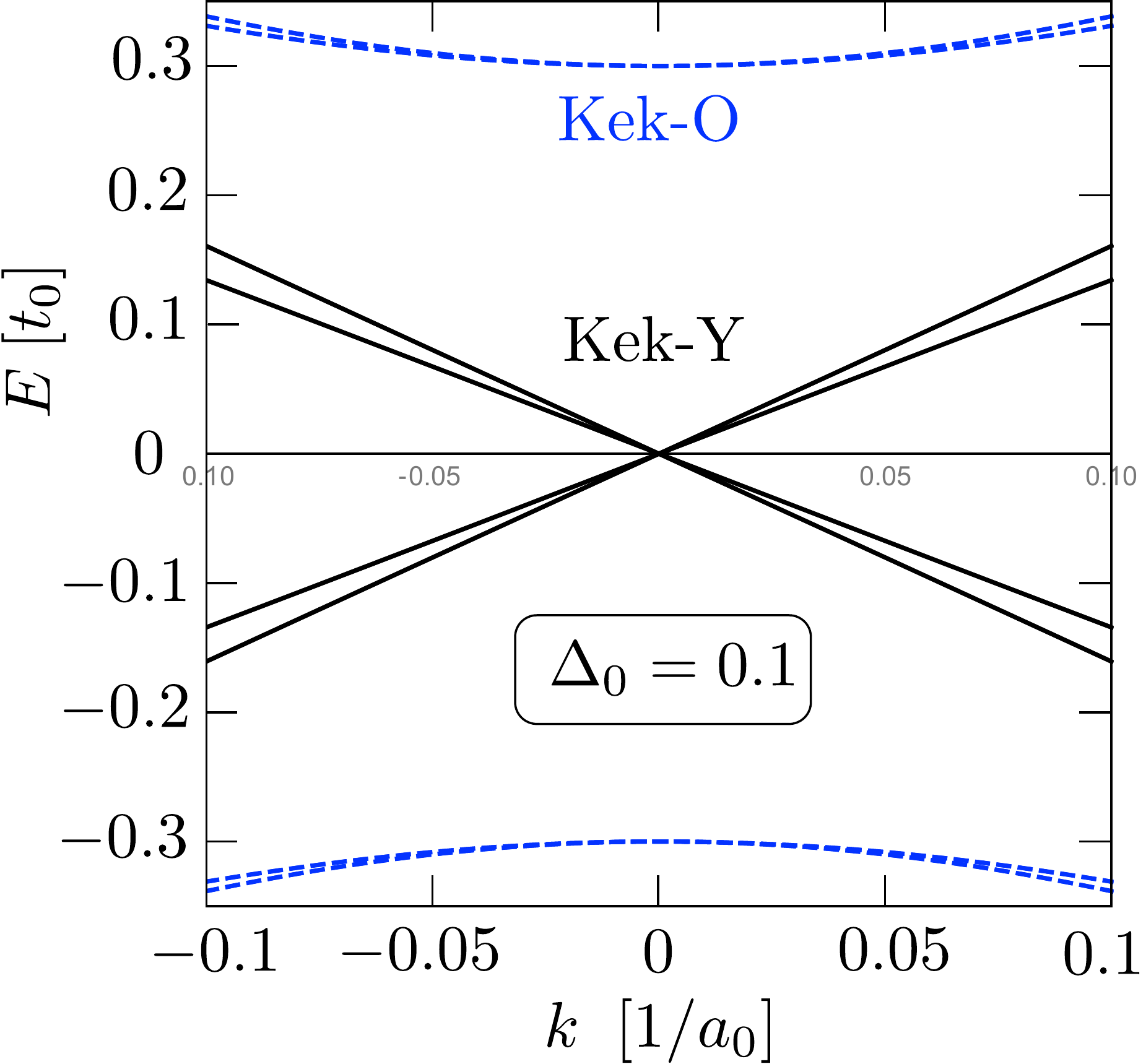}}
\caption{Dispersion relation near the center of the superlattice Brillouin zone, for the Kek-O texture (blue dashed curves) and for the Kek-Y texture (black solid). The curves are calculated from the full Hamiltonian \eqref{Hkreduceddef} for $|\tilde{\Delta}|=\Delta_0=0.1$.
}
\label{fig_bandstructure}
\end{figure}

For the Kek-O texture we recover the gapped spectrum of Kekul\'{e} dimerized graphene \cite{Cha00},
\begin{equation}
E^2=v_0^2 |\bm{p}|^2+(3t_0\Delta_0)^2\;\;\text{for}\;\;\nu=0.\label{EKekO}
\end{equation}
The Kek-Y texture, instead, has a gapless spectrum,
\begin{equation}
E_\pm^2=v_0^2(1\pm\Delta_0)^2|\bm{p}|^2,\;\;\text{for}\;\;|\nu|=1,\label{EKekY}
\end{equation}
consisting of a pair of linearly dispersing modes with different velocities $v_0(1\pm\Delta_0)$. The two qualitatively different dispersions are contrasted in Fig.\ \ref{fig_bandstructure}.

\subsection{Valley-momentum locking}
\label{valleymomentumlock}

The two gapless modes in the Kek-Y superlattice are helical, with both the sublattice pseudospin and the valley isospin locked to the direction of motion. To see this, we consider the $\nu=1$ Kek-Y texture with a real $\tilde{\Delta}=\Delta_0$. (Complex $\tilde{\Delta}$ and $\nu=-1$ are equivalent upon a unitary transformation.) The Dirac Hamiltonian \eqref{HDiracdef} can be written in the compact form
\begin{equation}
{\cal H}=v_\sigma\,(\bm{p}\cdot\bm{\sigma})\otimes\tau_0+v_\tau\,\sigma_0\otimes(\bm{p}\cdot\bm{\tau}),\label{calHdef}
\end{equation}
with the help of a  second set of Pauli matrices $\tau_x,\tau_y,\tau_z$ and unit matrix $\tau_0$ acting on the valley degree of freedom. The two velocities are defined by $v_\sigma=v_0$ and $v_\tau=v_0\Delta_0$.

An eigenstate of the current operator
\begin{equation}
j_\alpha=\partial{\cal H}/\partial p_\alpha=v_\sigma\,\sigma_\alpha\otimes\tau_0+v_\tau\,\sigma_0\otimes\tau_\alpha\label{jalphadef}
\end{equation}
with eigenvalue $v_\sigma\pm v_\tau$ is an eigenstate of $\sigma_\alpha$ with eigenvalue $+1$ and an eigenstate of $\tau_\alpha$ with eigenvalue $\pm 1$. (The two Pauli matrices act on different degrees of freedom, so they commute and can be diagonalized independently.) This valley-momentum locking does not violate time-reversal symmetry, since the time-reversal operation in the superlattice inverts all three vectors $\bm{p}$, $\bm{\sigma}$, and $\bm{\tau}$, and hence leaves $\cal H$ unaffected \cite{note2}:
\begin{equation}
(\sigma_y\otimes\tau_y){\cal H}^\ast(\sigma_y\otimes\tau_y)={\cal H}.\label{TRS}
\end{equation}

The valley-momentum locking does break the sublattice symmetry, since ${\cal H}$ no longer anticommutes with $\sigma_z$, but another chiral symmetry involving both sublattice and valley degrees of freedom remains:
\begin{equation}
(\sigma_z\otimes\tau_z){\cal H}=-{\cal H}(\sigma_z\otimes\tau_z).\label{chiralsymm}
\end{equation}

\subsection{Landau level quantization}
\label{Landaulevelquant}

A perpendicular magnetic field $B$ in the $z$-direction (vector potential $\bm{A}$ in the $x$--$y$ plane), breaks the time-reversal symmetry \eqref{TRS} via the substitution $\bm{p}\mapsto-i\hbar\partial/\partial\bm{r}+e\bm{A}(\bm{r})\equiv\bm{\Pi}$. The chiral symmetry \eqref{chiralsymm} is preserved, so the Landau levels are still symmetrically arranged around $E=0$, as in unperturbed graphene. Because the two helicity operators $\bm{\Pi}\cdot\bm{\sigma}$ and $\bm{\Pi}\cdot\bm{\tau}$ do not commute for $\bm{A}\neq 0$, they can no longer be diagonalized independently. In particular, this means the Landau level spectrum is not simply a superposition of two spectra of Dirac fermions with different velocities.

It is still possible to calculate the spectrum analytically (see App.\ \ref{appA}). We find Landau levels at energies $E_{n}^+,E_{n}^-,-E_{n}^+,-E_{n}^-$, $n=0,1,2,\ldots$, given by
\begin{equation}
E_n^\pm=E_B\left[2n+1\pm
 \sqrt{1+n(n+1)(4v_\sigma v_\tau)^2\bar{v}^{-4} }\right]^{1/2},\label{Enpmresult}
\end{equation}
with the definitions $\bar{v}=\sqrt{v_\sigma^2+ v_\tau^2}$ and $E_B=\bar{v}\sqrt{\hbar eB}$. 

\begin{figure}[tb]
\centerline{\includegraphics[width=0.7\linewidth]{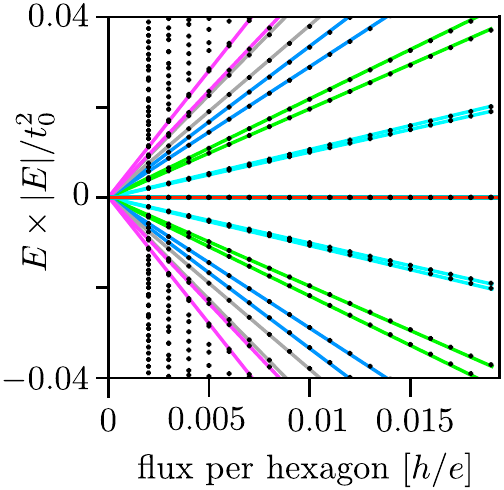}}
\caption{Landau levels in the Kek-Y superlattice ($\Delta_0=0.1$, $\phi=0$, $\nu=1$). The data points are calculated numerically \cite{kwant} from the tight-binding Hamiltonian \eqref{Ham1} with bond modulation \eqref{trndef}. The lines are the analytical result from Eqs.\ \eqref{Enpmresult} and \eqref{rhopmresult} for the first few Landau levels. Lines of the same color identify the valley-split Landau level, the zeroth Landau level (red line) is not split.
}
\label{fig_landaufan}
\end{figure}

In unperturbed graphene all Landau levels have a twofold valley degeneracy \cite{note3}: $E_{n}^+=E_{n+1}^-$ for $v_\tau=0$. This includes the zeroth Landau level: $E_{0}^-=0=-E_{0}^-$. A nonzero $v_\tau$ breaks the valley degeneracy of all Landau levels at $E\neq 0$, but a valley-degenerate zero-mode $E_0^-=0$ remains, see Fig.\ \ref{fig_landaufan}.

The absence of a splitting of the zeroth-Landau level can be understood as a topological protection in the context of an index theorem \cite{Aha79,Wen89,Kat08,Kai09}, which requires that \textit{either} $\Pi_+\equiv\Pi_x+i\Pi_y$ or $\Pi_-\equiv\Pi_x-i\Pi_y$ has a zero-mode. If we decompose ${\cal H}=\Pi_+ S_- +\Pi_- S_+$, with $S_\pm=v_\sigma(\sigma_x\pm i\sigma_y)+v_\tau(\tau_x\pm i\tau_y)$, we see that \textit{both} $S_+$ and $S_-$ have a rank-two null space \cite{note4}, spanned by the spinors $\psi_\pm^{(1)}$ and $\psi_\pm^{(2)}$. So if $\Pi_\pm f_\pm=0$, a twofold degenerate zero-mode of ${\cal H}$ is formed by the states $f_\pm\psi_\mp^{(1)}$ and $f_\pm\psi_\mp^{(2)}$.

All of this is distinctive for the Kek-Y bond order: for the Kek-O texture it's the other way around --- the Landau levels have a twofold valley degeneracy except for the nondegenerate Landau level at the edge of the band gap \cite{note6}.

\section{Effect of virtual transitions to higher bands}
\label{virtualtransitions}

So far we have assumed $\Delta_0\ll 1$, and one might ask how robust our findings are to finite-$\Delta_0$ corrections, involving virtual transitions from the $\varepsilon_{\pm 1}$ bands near $E=0$ to the $\varepsilon_0$ band near $E=3t_0$. We have been able to include these to all orders in $\Delta_0$ (see App.\ \ref{appB}), and find that the entire effect is a renormalization of the velocities $v_\sigma$ and $v_\tau$ in the Hamiltonian \eqref{calHdef}, which retains its form as a sum of two helicity operators. For real $\Delta=\Delta_0$ the renormalization is given by $v_\sigma=v_0\rho_+$, $v_\tau=v_0\rho_-$ with
\begin{equation}
\rho_\pm=\tfrac{1}{2}(1-\Delta_0)\left(\frac{1+2\Delta_0}{\sqrt{1+2\Delta_0^2}}\pm 1\right).\label{rhopmresult}
\end{equation}
For complex $\Delta=\Delta_0 e^{i\phi}$ the nonlinear renormalization introduces a dependence on the phase $\phi$ modulo $2\pi/3$. 

What this renormalization shows is that, as expected for a topological protection, the robustness of the zeroth Landau level to the Kek-Y texture is not limited to perturbation theory --- also strong modulations of the bond strength cannot split it away from $E=0$.

\section{Pseudospin-valley coupling}
\label{sigmataucoupling}

In zero magnetic field the low-energy Hamiltonian \eqref{calHdef} does not couple the pseudospin $\sigma$ and valley $\tau$ degrees of freedom. A $\bm{\sigma}\otimes\bm{\tau}$ coupling is introduced in the Kek-Y superlattice by an ionic potential $\mu_{\rm Y}$ on the carbon atoms that line up with the carbon vacancies --- the atoms located at each center of a red Y in Fig.\ \ref{FLattice}. We consider this effect for the $\nu=1$ Kek-Y texture with a real $\tilde{\Delta}=\Delta_0$. 

The ionic potential acts on one-third of the A sublattice sites, labeled $\bm{r}_{\rm Y}$. (For $\nu=-1$ it would act on one-third of the B sublattice sites.) Fourier transformation of the on-site contribution $\mu_{\rm Y}\sum_{\bm{r}_{\rm Y}}a^\dagger_{\bm{r}_{\rm Y}}a_{\bm{r}_{\rm Y}}$ to the tight-binding Hamiltonian \eqref{Ham1} gives with the help of the lattice sum
\begin{equation}
\textstyle{\sum_{\bm{r}_{\rm Y}}}e^{i\bm{k}\cdot\bm{r}_{\rm Y}}\propto\delta(\bm{k})+\delta(\bm{k}-\bm{G})+\delta(\bm{k}+\bm{G})\label{rYlatticesum}
\end{equation}
the momentum-space Hamiltonian
\begin{subequations}
\label{HMdef}
\begin{align}
&H(\bm{k})=-c^\dagger_{\bm{k}}\begin{pmatrix}
M_{\rm Y}&{\cal E}_{1}(\bm{k})\\
{\cal E}^\dagger_{1}(\bm{k})&0
\end{pmatrix}c_{\bm k},\label{HMdefa}\\
&M_{\rm Y}=-\mu_{\rm Y}\begin{pmatrix}
1&1&1\\
1&1&1\\
1&1&1
\end{pmatrix}.\label{HMdefb}
\end{align}
\end{subequations}
The ${\cal E}_1$ block is still given by Eq.\ \eqref{Hkreduceddef}. The additional $M_{\rm Y}$-block breaks the chiral symmetry.

Projection onto the subspace spanned by $u_{\bm k}=(a_{\bm{k}-\bm{G}},a_{\bm{k}+\bm{G}},b_{\bm{k}-\bm{G}},b_{\bm{k}+\bm{G}})$ gives the effective Hamiltonian
\begin{equation}
H_{\rm eff}=-u^\dagger_{\bm{k}}\begin{pmatrix}
m_{\rm Y}&h_1\\
h^\dagger_1&0
\end{pmatrix}u_{\bm k},\;\;m_{\rm Y}=-\mu_{\rm Y}\begin{pmatrix}
1&1\\
1&1
\end{pmatrix}.\label{Hmeff}
\end{equation}
The corresponding Dirac Hamiltonian has the form \eqref{HDiracdef} with an additional $\bm{\sigma}\otimes\bm{\tau}$ coupling,
\begin{equation}
\begin{split}
{\cal H}={}&v_\sigma\,(\bm{p}\cdot\bm{\sigma})\otimes\tau_0+v_\tau\,\sigma_0\otimes(\bm{p}\cdot\bm{\tau})+\tfrac{1}{2}\mu_{\rm Y}\\
&+\tfrac{1}{2}\mu_{\rm Y}(\sigma_x\otimes\tau_x+\sigma_y\otimes\tau_y-\sigma_z\otimes\tau_z).
\end{split}
\label{calHmdef}
\end{equation}

\begin{figure}[tb]
\centerline{\includegraphics[width=0.7\linewidth]{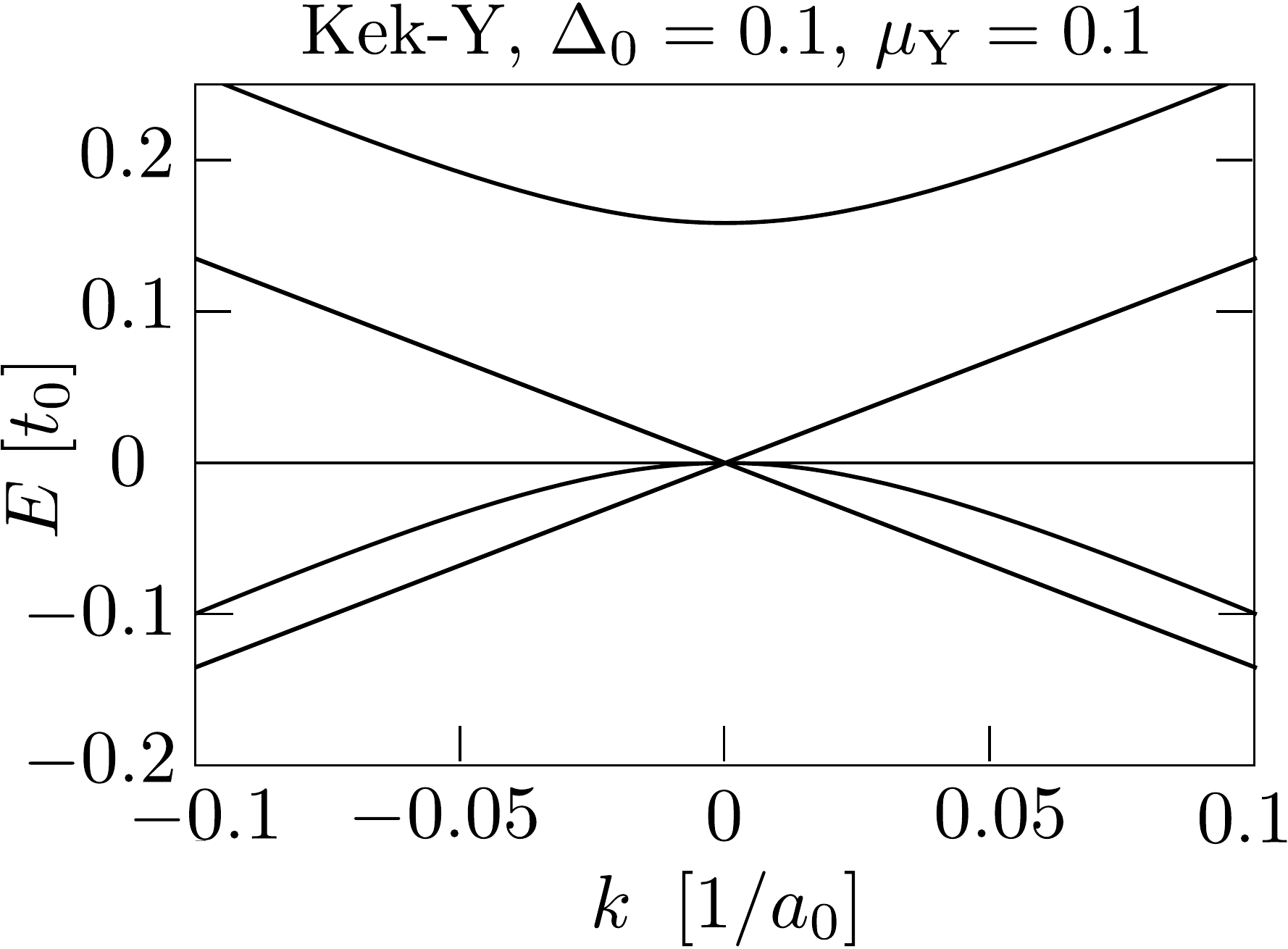}}
\caption{Effect of an on-site potential $\mu_{\rm Y}$ on the Kek-Y bandstructure of Fig.\ \ref{fig_bandstructure}. The three bands that intersect linearly and quadratically at the center of the superlattice Brillouin zone form the ``spin-one Dirac cone'' of Refs.\ \onlinecite{Gio15} and \onlinecite{Ren15}. The curves are calculated from the full Hamiltonian \eqref{HMdef} for $\Delta_0=0.1=\mu_{\rm Y}$.
}
\label{fig_bandstructure_2}
\end{figure}

The energy spectrum,
\begin{equation}
\begin{split}
&E_\pm^{(1)} = \pm (v_\sigma-v_\tau)|\bm{p}|,\\
& E_\pm^{(2)} = \mu_{\rm Y} \pm \sqrt{(v_\sigma+v_\tau)^2|\bm{p}|^2+\mu_{\rm Y}^2},
\end{split}
\label{fourbands}
\end{equation}
has two bands that cross linearly in $p$ at $E=0$, while the other two bands have a quadratic $p$-dependence. (See Fig.\ \ref{fig_bandstructure_2}.)

The three bands $E_{+}^{(1)}$, $E_{-}^{(1)}$, $E_-^{(2)}$ that intersect at $p=0$ are reminiscent of a spin-one Dirac one. Such a dispersion is a known feature of a potential modulation that involves only one-third of the atoms on one sublattice \cite{Gio15,Ren15}. The spectrum remains gapless even though the chiral symmetry is broken. This is in contrast to the usual staggered potential between A and B sublattices, which opens a gap via a $\sigma_z\otimes\tau_z$ term \cite{Cas09}.

\section{Discussion}
\label{discuss}

In summary, we have shown that the Y-shaped Kekul\'{e} bond texture (Kek-Y superlattice) in graphene preserves the massless character of the Dirac fermions. This is fundamentally different from the gapped band structure resulting from the original Kekul\'{e} dimerization \cite{Cha00,Hou07,Che09a,Che09b} (Kek-O superlattice), and contrary to expectations from its experimental realization \cite{Gut16,Gut15}.

The gapless low-energy Hamiltonian ${\cal H}=v_\sigma\bm{p}\cdot\bm{\sigma}+v_\tau\bm{p}\cdot\bm{\tau}$ is the sum of two helicity operators, with the momentum $\bm{p}$ coupled independently to both the sublattice pseudospin $\bm{\sigma}$ and the valley isospin $\bm{\tau}$. This valley-momentum locking is distinct from the coupling of the valley to a pseudo-magnetic field that has been explored as an enabler for valleytronics \cite{Wan15}, and offers a way for a momentum-controlled valley precession. The broken valley degeneracy would also remove a major obstacle for spin qubits in graphene \cite{Tra07}.

A key experimental test of our theoretical predictions would be a confirmation that the Kek-Y superlattice has a gapless spectrum, in stark contrast to the gapped Kek-O spectrum. In the experiment by Guti\'{e}rrez \textit{et al.}\ on a graphene/Cu heterostructure the Kek-Y superlattice is formed by copper vacancies that are in registry with one out of six carbon atoms \cite{Gut16,Gut15}. These introduce the Y-shaped hopping modulations shown in Fig.\ \ref{FLattice}, but in addition will modify the ionic potential felt by the carbon atom at the center of the Y. Unlike the usual staggered potential between A and B sublattices, this potential modulation in an enlarged unit cell does not open a gap \cite{Gio15,Ren15}. We have also checked that the Dirac cone remains gapless if we include hoppings beyond nearest neigbor. All of this gives confidence that the gapless spectrum will survive in a realistic situation.

Further research in other directions could involve the Landau level spectrum, to search for the unique feature of a broken valley degeneracy coexisting with a valley-degenerate zero-mode. The graphene analogues in optics and acoustics \cite{Pol13} could also provide an interesting platform for a Kek-Y superlattice with a much stronger amplitude modulation than can be realized with electrons.

\acknowledgments

We have benefited from discussions with A. Akhmerov, V. Cheianov, J. Hutasoit, P. Silvestrov, and D. Varjas. This research was supported by the Netherlands Organization for Scientific Research (NWO/OCW) and an ERC Synergy Grant.

\appendix

\section{Calculation of the Landau level spectrum in a Kek-Y superlattice}
\label{appA}

We calculate the spectrum in a perpendicular magnetic field of a graphene sheet with a Kekul\'{e}-Y bond texture. We start by rewriting the Hamiltonian \eqref{calHdef}, with $\bm{\Pi}=\bm{p}+e\bm{A}$, in the form
\begin{equation}
{\cal H}=\tfrac{1}{2}\Pi_+ S_- +\tfrac{1}{2}\Pi_- S_+ +\mu\sigma_z\otimes\tau_z,\label{HPiS}
\end{equation}
in terms of the raising and lowering operators
\begin{equation}
\begin{split}
&\Pi_\pm=\Pi_x\pm i\Pi_y,\;\;\sigma_\pm=\sigma_x\pm i\sigma_y,\;\;\tau_\pm=\tau_x\pm i\tau_y,\\
& S_\pm=v_\sigma\,\sigma_\pm\otimes\tau_0+v_\tau\,\sigma_0\otimes\tau_\pm.
\end{split}
\label{raisinglowering}
\end{equation}
The chiral-symmetry breaking term $\mu\sigma_z\otimes\tau_z$ that we have added will serve a purpose later on.

We know that the Hermitian operator $\Omega=\Pi_+\Pi_-$ has eigenvalues $\omega_n=2n\hbar eB$, $n=0,1,2,\ldots$, in view of the commutator $[\Pi_-,\Pi_+]=2\hbar eB$. So the strategy is to express the secular equation $\det(E-{\cal H})=0$ in a form that involves only the mixed products $\Pi_+\Pi_-$, and no $\Pi_+^2$ or $\Pi_-^2$. This is achieved by means of a unitary transformation, as follows. 

We define the unitary matrix
\begin{equation}
U=\exp[\tfrac{1}{4}i\pi(\sigma_0+\sigma_z)\otimes\tau_y]\label{Uunitarydef}
\end{equation}
and reduce the determinant of a $4\times 4$ matrix to that of a $2\times 2$ matrix:
\begin{align}
&\det({\cal H}-E)=\det U^\dagger({\cal H}-E)U\nonumber\\
&\quad=\det\begin{pmatrix}
-E+\mu&R^\dagger\\
R&-E-\mu
\end{pmatrix}\nonumber\\
&\quad=\begin{cases}
\det(E^2-\mu^2-RR^\dagger )&{\rm if}\;\;E\neq \mu,\\
\det(E^2-\mu^2-R^\dagger R)&{\rm if}\;\;E\neq-\mu,
\end{cases}\\
&\qquad\text{with}\;\; R=\begin{pmatrix}
-v_\tau\Pi_-&v_\sigma\Pi_-\\
-v_\sigma\Pi_+& v_\tau\Pi_+
\end{pmatrix}.\label{Rdef}
\end{align}
\begin{widetext}
The matrix product $RR^\dagger$ is not of the desired form, but $R^\dagger R$ is, 
\begin{equation}
R^\dagger R=\begin{pmatrix}
v_\sigma^2\Pi_-\Pi_+ + v_\tau^2\Pi_+\Pi_- & -v_\sigma v_\tau(\Pi_-\Pi_+ +\Pi_+\Pi_-)\\
-v_\sigma v_\tau(\Pi_-\Pi_+ +\Pi_+\Pi_-) & v_\sigma^2\Pi_+\Pi_- + v_\tau^2\Pi_-\Pi_+
\end{pmatrix},\label{RRdagger}
\end{equation}
involving only $\Pi_+\Pi_-=\Omega$ and $\Pi_-\Pi_+=\Omega+\omega_1$. Hence the determinant is readily evaluated for $E\neq -\mu$,
\begin{equation}
\det({\cal H}-E)=\det(E^2-\mu^2-R^\dagger R)=\prod_{n=0}^\infty\det\begin{pmatrix}
E^2-\mu^2-\bar{v}^2\omega_n-v_\sigma^2\omega_1&v_\sigma v_\tau(2\omega_n+\omega_1)\\
v_\sigma v_\tau(2\omega_n+\omega_1)&E^{2}-\mu^2-\bar{v}^2\omega_n- v_\tau^2\omega_1
\end{pmatrix},
\end{equation}
where we have abbreviated $\bar{v}=\sqrt{v_\sigma^2+ v_\tau^2}$.

Equating the determinant to zero and solving for $E$ we find four sets of energy eigenvalues $E_{n}^+,E_{n}^-,-E_{n}^+,-E_{n}^-$, given by
\begin{equation}
(E_n^\pm)^2-\mu^2=(\omega_n+\tfrac{1}{2}\omega_1)\bar{v}^2\pm \tfrac{1}{2}\sqrt{\omega_1^2\bar{v}^4+(4v_\sigma v_\tau)^2\omega_n\omega_{n+1}}=E_B^2\left[2n+1\pm
 \sqrt{1+n(n+1)(4v_\sigma v_\tau)^2\bar{v}^{-4}} \right].\label{Enappresult}
\end{equation}
In the second equation we introduced the energy scale $E_B=\hbar\bar{v}/l_m$, with $l_m=\sqrt{\hbar/eB}$ the magnetic length. The $B$-independent level $E_0^-=\mu$ becomes a zero-mode in the limit $\mu\rightarrow 0$.
\end{widetext}

As a check on the calculation, we note that for $\mu=0$, $v_\tau=0$ we recover the valley-degenerate Landau level spectrum of graphene \cite{Cas09},
\begin{equation}
E_{n}^-=(\hbar v_\sigma/l_m)\sqrt{2n},\;\; E_n^+=E_{n+1}^-.\label{Edeltavis0}
\end{equation}

Another special case of interest is $\mu=0$, $v_\sigma=v_\tau\equiv v_0$, when the two modes of Dirac fermions have velocities $v_\sigma\pm v_\tau$ equal to $0$ and $2v_0$. From Eq.\ \eqref{Enappresult}  we find the Landau level spectrum
\begin{equation}
E_n^-=0,\;\;E_n^+= 2(\hbar v_0/l_m)\sqrt{2n+1}.\label{Edeltavisv}
\end{equation}
The mode with zero velocity remains $B$-independent, while the mode with velocity $2v_0$ produces a sequence of Landau levels with a $1/2$ offset in the $n$-dependence.

\section{Calculation of the low-energy Hamiltonian to all orders in the Kek-Y bond modulation}
\label{appB}

\begin{figure}[tb]
\centerline{\includegraphics[width=0.8\linewidth]{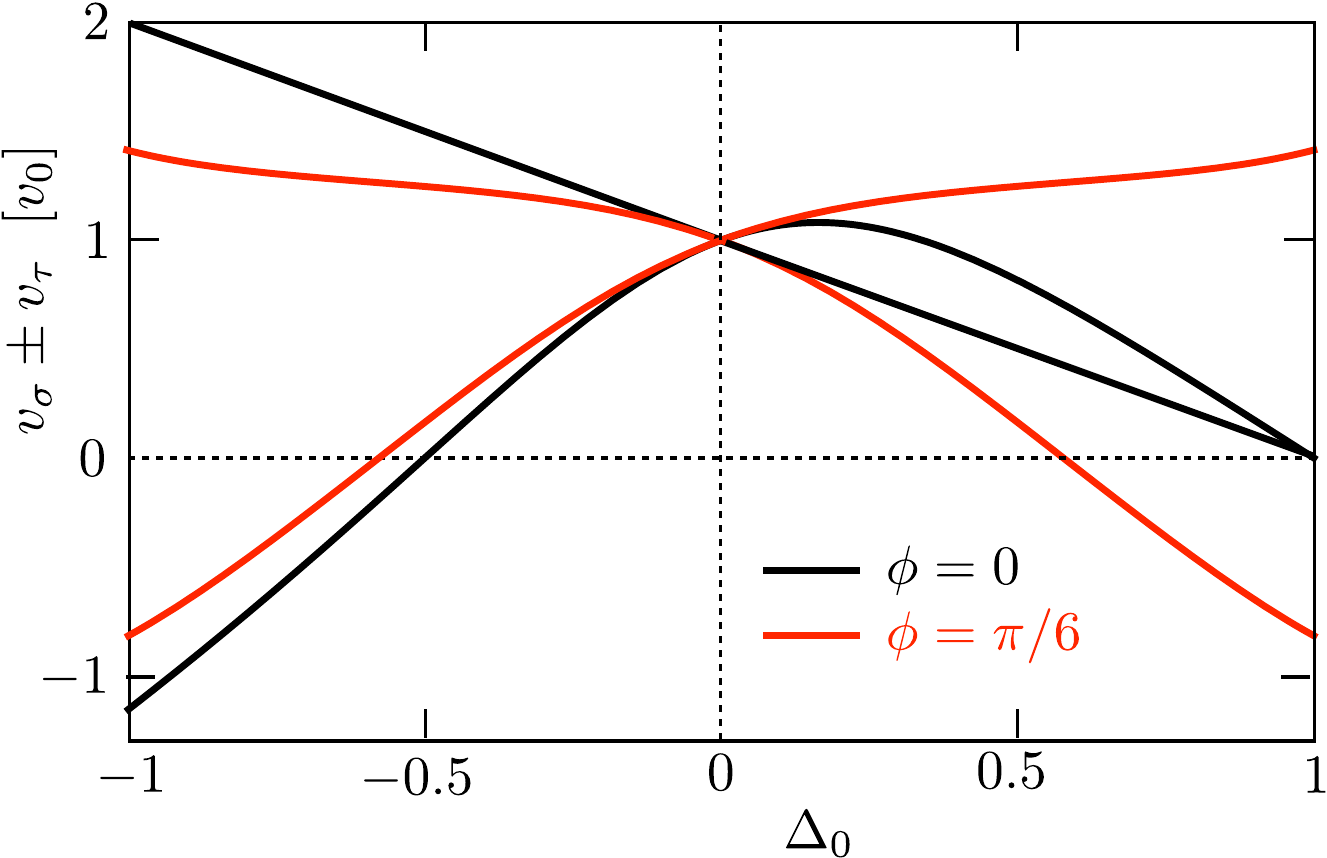}}
\caption{Velocities $v_1=v_\sigma+ v_\tau$ and $v_2=v_\sigma- v_\tau$ of the two gapless modes in the Kek-Y superlattice, as a function of the bond modulation amplitude $\Delta_0$ for two values of the modulation phase $\phi$. The $\phi$-dependence modulo $2\pi/3$ appears to second order in $\Delta_0$. The curves are calculated from Eq.\ \eqref{absrhopm}. Note that positive and negative values of $v_1,v_2$ are equivalent.
}
\label{fig_velocities}
\end{figure}

\begin{figure}[tb]
\centerline{\includegraphics[width=0.6\linewidth]{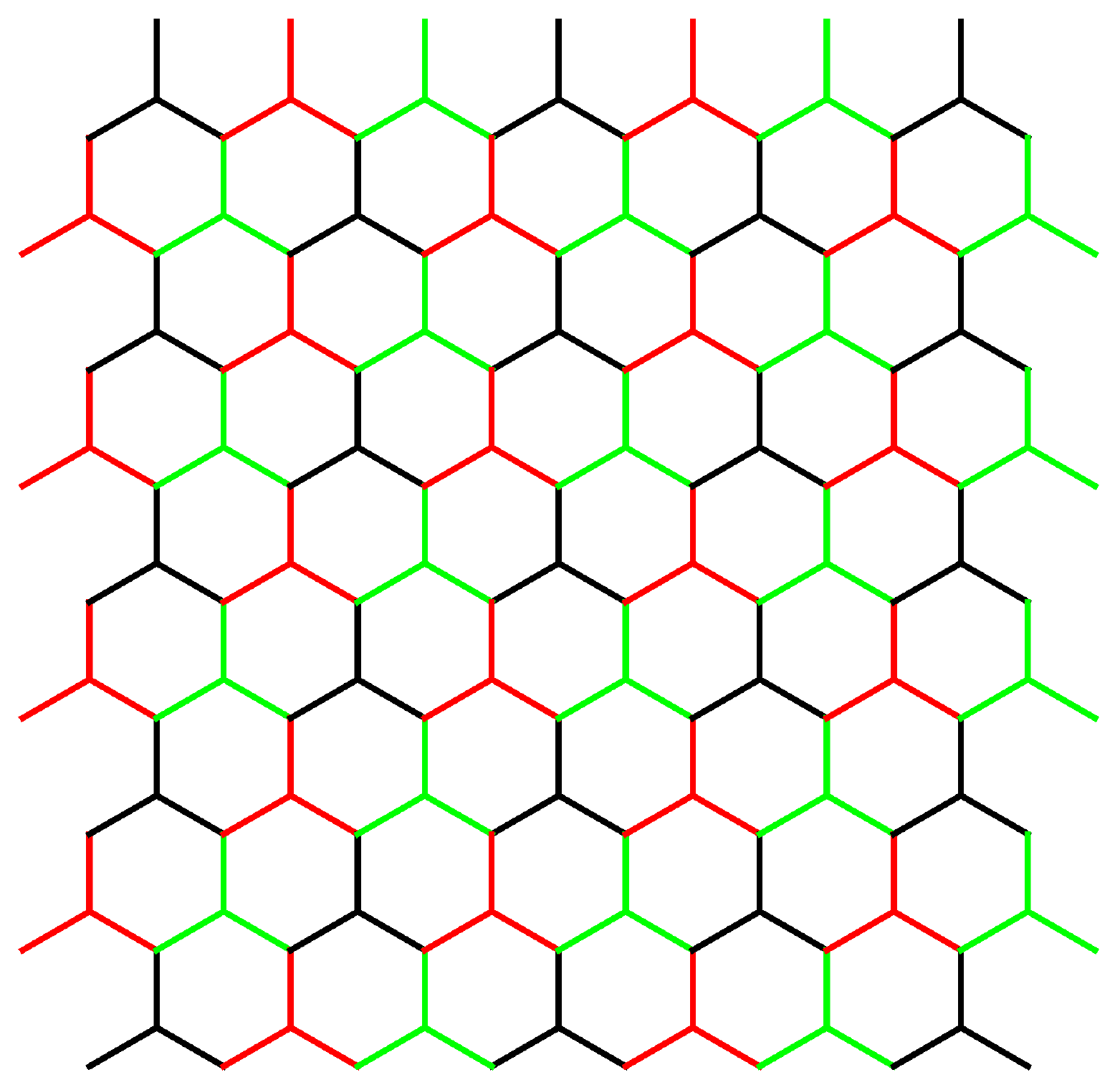}}
\caption{Kek-Y superlattice with a complex bond amplitude $\Delta=e^{i\phi}\Delta_0$, according to Eq.\ \eqref{trndef} with $\nu=1$. The three colors of the bonds refer to three different bond strengths, adding up to $3t_0$. For $\phi=0$  two of the bond strengths are equal to $t_0(1-\Delta_0)$ and the third equals $t_0(1+2\Delta_0)$. This is the case shown in Fig.\ \ref{FLattice}. For $\phi=\pi/6$ the bond strengths are equidistant: $t_0(1-\Delta_0\sqrt 3)$, $t_0$, and $t_0(1+\Delta_0\sqrt 3)$. The value of $\Delta_0$ where a bond strength vanishes shows up in Fig.\ \ref{fig_velocities} as a point of vanishing velocity.
}
\label{symmetricKeky}
\end{figure}

We seek to reduce the six-band Hamiltonian \eqref{Hkreduceddef} to an effective $4\times 4$ Hamiltonian that describes the low-energy spectrum near $\bm{k}=0$. For $\Delta_0\ll 1$ we can simply project onto the $2\times 2$ lower-right subblock of ${\cal E}_\nu$, which for the $|\nu|=1$ Kek-Y bond modulation vanishes linearly in $\bm{k}$. This subblock is coupled to the $\varepsilon_0$ band near $E=3t_0$ by matrix elements of order $\Delta_0$, so virtual transitions to this higher band contribute to the low-energy spectrum in order $\Delta_0^2$. We will now show how to include these effects to all order in $\Delta_0$.

One complication when we go beyond the small-$\Delta_0$ regime is that the phase $\phi$ of the modulation amplitude can no longer be removed by a unitary transformation. As we will see, the low-energy Hamiltonian depends on $\phi$ modulo $2\pi/3$ --- so we don't need to distinguish between the phase of $\tilde{\Delta}=e^{2\pi i(p+q)/3}\Delta$ and the phase of $\Delta$. The choice between $\nu=\pm 1$ still does not matter, the two Kek-Y modulations being related by a mirror symmetry. For definiteness we take $\nu=+1$.

We define the unitary matrix
\begin{subequations}
\begin{align}
&V=\begin{pmatrix}
\Phi&0\\
0&\Phi
\end{pmatrix}\begin{pmatrix}
{\cal V}&0\\
0&\openone
\end{pmatrix},\;\;
\Phi=\begin{pmatrix}
1&0&0\\
0&e^{-i\phi}&0\\
0&0&e^{i\phi}
\end{pmatrix},\\
&{\cal V}=\frac{1}{2D_0}
\begin{pmatrix}
2 & -2\Delta_0 & -2\Delta_0\\
2\Delta_0& 1+D_0 & 1-D_0\\
 2\Delta_0& 1-D_0 & 1+D_0\end{pmatrix},\label{VcalVdef}
\end{align}
\end{subequations}
with $D_0=\sqrt{1+2\Delta_0^2}$ and evaluate
\begin{subequations}
\begin{align}
&V^\dagger\begin{pmatrix}
0&{\cal E}_1\\
{\cal E}^\dagger_1&0
\end{pmatrix}V=\begin{pmatrix}
0&\tilde{\cal E}_1\\
\tilde{\cal E}_1^\dagger&0
\end{pmatrix},\\
&\tilde{\cal E}_1={\cal V}^\dagger{\cal E}_1=
\begin{pmatrix}
D_0\varepsilon_0&\rho_0^\ast\varepsilon_{-1}&\rho_0\varepsilon_{1}\\
0&\rho_+\varepsilon_{-1}&\rho_-^\ast\varepsilon_1\\
0&\rho_-\varepsilon_{-1}&\rho_+^\ast\varepsilon_1
\end{pmatrix},\\
&\rho_\pm=\frac{1}{2D_0}\left[1-2\Delta_0^2\pm D_0+e^{-3i\phi}\Delta_0(1\mp D_0)\right],\\
&\rho_0=\frac{\Delta_0}{D_0}(2+e^{3i\phi}\Delta_0).
\end{align}
\end{subequations}
The matrix elements that couple the lower-right $2\times 2$ subblock of $\tilde{\cal E}_1$ to $\varepsilon_0$ are now of order $k$, so the effect on the low-energy spectrum is of order $k^2$ and can be neglected --- \textit{to all orders in $\Delta_0$}.

The resulting effective low-energy Hamiltonian has the $4\times 4$ form \eqref{Hkeff}, with $h_{1}$ replaced by 
\begin{equation}
{h}_1=\begin{pmatrix}
\rho_+\varepsilon_{-1}&\rho_-^\ast\varepsilon_1\\
\rho_-\varepsilon_{-1}&\rho_+^\ast\varepsilon_1
\end{pmatrix}.\label{Hkeff2}
\end{equation}
The phases of $\rho_\pm=|\rho_\pm|e^{i\theta_\pm}$ can be eliminated by one more unitary transformation, with the $4\times 4$ diagonal matrix
\begin{equation}
\Theta={\rm diag}\,(e^{i\theta_-},e^{i\theta_+},e^{i\theta_++i\theta_-},1),\label{Thetadef}
\end{equation}
which results in
\begin{equation}
\Theta^\dagger\begin{pmatrix}
0&{h}_1\\
\tilde{h}_1^\dagger&0
\end{pmatrix}\Theta
=\begin{pmatrix}
0&\tilde{h}_1\\
\tilde{h}_1^\dagger&0
\end{pmatrix},\;\;\tilde{h}_1=\begin{pmatrix}
|\rho_+|\varepsilon_{-1}&|\rho_-|\varepsilon_1\\
|\rho_-|\varepsilon_{-1}&|\rho_+|\varepsilon_1
\end{pmatrix}.
\end{equation}

Finally, we arrive at the effective Hamiltonian \eqref{calHdef}, with renormalized velocities:
\begin{widetext}
\begin{align}
&{\cal H}=v_\sigma\,(\bm{p}\cdot\bm{\sigma})\otimes\tau_0+v_\tau\,\sigma_0\otimes(\bm{p}\cdot\bm{\tau}),\;\;v_\sigma=|\rho_+|v_0,\;\;v_\tau=|\rho_-|v_0,\\
&|\rho_\pm|^2=\frac{1}{2D_0^2}\biggl(1+3\Delta_0^4\pm D_0(1- 3\Delta_0^2) +2\Delta_0^3(\pm D_0-2)\cos 3\phi\biggr).\label{absrhopm}
\end{align}
To third order in $\Delta_0$ we have
\begin{equation}
v_\sigma/v_0=1-\tfrac{3}{2}\Delta_0^2-\tfrac{1}{2}\Delta_0^3\cos 3\phi,\;\;v_\tau/v_0=\Delta_0-\tfrac{3}{2}\Delta_0^2\cos 3\phi+\tfrac{1}{16}\Delta_0^3(1-9\cos 6\phi)+{\cal O}(\Delta_0^4).
\end{equation}
\end{widetext}

For real $\Delta$, when $\phi=0$ and $\rho_\pm$ is real, Eq.\ \eqref{absrhopm} simplifies to
\begin{equation}
\rho_\pm=\tfrac{1}{2}(1-\Delta_0)\left(\frac{1+2\Delta_0}{\sqrt{1+2\Delta_0^2}}\pm 1\right).
\end{equation}
The velocities of the two Dirac modes are then given by
\begin{equation}
\begin{split}
&v_1=v_\sigma+v_\tau=v_0\frac{(1-\Delta_0)(1+2\Delta_0)}{\sqrt{1+2\Delta_0^2}}\\
&v_2=v_\sigma-v_\tau=v_0(1-\Delta_0).
\end{split}
\end{equation}
More generally, for complex $\Delta=\Delta_0 e^{i\phi}$ both $v_1$ and $v_2$ become $\phi$-dependent to second order in $\Delta_0$, see Fig.\ \ref{fig_velocities}. 

Note that the asymmetry in $\pm\Delta_0$ vanishes for $\phi=\pi/6$. For this phase the superlattice has three different bond strengths (see Fig.\ \ref{symmetricKeky}) that are symmetrically arranged around the unperturbed value $t_0$. 
 

\begin{thebibliography}{99}
\bibitem{note0} There are three sets of integers $p,q\in\mathbb{Z}_3$ for a given index $\nu=1+q-p\mod 3$, corresponding to textures on the honeycomb lattice that are translated by one hexagon, or equivalently related by a $\pm 2\pi/3$ phase shift of $\Delta$.
\bibitem{note2} The time-reversal operation ${\cal T}=(\sigma_y\otimes\tau_y){\cal C}$ from Eq.\ \eqref{TRS} (with ${\cal C}$ complex conjugation) squares to $+1$ because the electron spin is not explicitly included. If we do include it, we would have ${\cal T}=(s_y\otimes\sigma_y\otimes\tau_y){\cal C}$, which squares to $-1$ as expected for a fermionic quasiparticle. The combination of the time-reversal symmetry \eqref{TRS} and the chiral symmetry \eqref{chiralsymm} places the superlattice in the BDI symmetry classification of topological states of matter.
\bibitem{note3} The Landau levels also have a twofold spin degeneracy, which could be resolved by the Zeeman energy but is not considered here.
\bibitem{note4} If we define the eigenstates $|\alpha,\beta\rangle$ by $\sigma_z|\alpha,\beta\rangle=\alpha|\alpha,\beta\rangle$, $\tau_z|\alpha,\beta\rangle=\beta|\alpha,\beta\rangle$, then $S_+$ annihilates $\psi_+^{(1)}=|1,1\rangle$ and $\psi_+^{(2)}=v_\tau|-1,1\rangle-v_\sigma|1,-1\rangle$, while $S_-$ annihilates $\psi_-^{(1)}=|-1,-1\rangle$ and $\psi_-^{(2)}=v_\tau|1,-1\rangle-v_\sigma|-1,1\rangle$.
\bibitem{note6} In a Kek-O superlattice the Landau levels are given by $E_{n}^2=(3t_0\Delta_0)^2+2n\hbar eBv_0^2$, $n=0,1,2,\ldots$, with a twofold valley degeneracy for $n\geq 1$ and a nondegenerate zeroth Landau level at $\pm 3t_0\Delta_0$.
\end{thebibliography}
\end{document}